\documentclass[aps, twocolumn, superscriptaddress]{revtex4} 

\usepackage[switch]{lineno}
\usepackage{amsmath}
\usepackage{amssymb}
\usepackage{empheq}
\usepackage[parfill]{parskip}
\usepackage[active]{srcltx}
\usepackage{color}
\usepackage{array}
\usepackage{booktabs}
\usepackage{amsfonts}
\usepackage{dsfont}
\usepackage{graphicx}
\usepackage{xcolor}

\begin{document}

\setlength{\parindent}{0.5cm}

\title{Swarmalators on a ring with distributed couplings}
\author{Kevin O'Keeffe}
\affiliation{Senseable City Lab, Massachusetts Institute of Technology, Cambridge, MA 02139} 
\email{Corresponding to: kevin.p.okeeffe@gmail.com}

\author{Hyunsuk Hong}
\affiliation{Department of Physics and Research Institute of Physics and Chemistry, Jeonbuk National University, Jeonju 54896, Korea} 
\email{Corresponding to: hhong@jbnu.ac.kr}
\affiliation{School of Physics, Korea Institute for Advanced Study, Seoul 02455, Korea} 

\begin{abstract}
We study a simple model of identical swarmalators, generalizations of phases oscillators that swarm through space. We confine the movements to a one-dimensional (1D) ring and consider distributed (non-identical) couplings; the combination of these two effects captures an aspect of the more realistic 2D swarmalator model \cite{o2017oscillators}. We find new collective states as well as generalizations of previously reported ones which we describe analytically. These states imitate the behavior of vinegar eels, catalytic microswimmers, and other swarmalators which move on quasi-1D rings.
\end{abstract}

\maketitle

\section{Introduction}
An interplay between sync \cite{winfree2001geometry, kuramoto2003chemical,pikovsky2003synchronization} (self-organization in time) and swarming \cite{bialek2012statistical, katz2011inferring} (self-organization in space) crops up everywhere in Nature, from biological microswimmers \cite{yang2008cooperation,riedel2005self,quillen2021metachronal,quillen2021synchronized,taylor1951analysis,tamm1975role} and chemical nanomotors \cite{yan2012linking,hwang2020cooperative,zhang2020reconfigurable,bricard2015emergent,zhang2021persistence,manna2021chemical,li2018spatiotemporal,chaudhary2014reconfigurable} to magnetic domains walls \cite{hrabec2018velocity,haltz2021domain} and robotic swarms \cite{barcis2019robots,barcis2020sandsbots,monaco2020cognitive}. Yet little is known about this dual form of self-organization from a theoretical perspective. Tanaka gave the first mathematical treatment of it  by deriving a model of chemotactic oscillators, oscillators which are pushed around by chemical gradients which in turn influence the oscillators' phases. \cite{tanaka2007general, iwasa2010dimensionality,iwasa2010hierarchical,iwasa2017mechanism}. Later O'Keeffe et al introduced a phenomenological model of `swarmlators` \cite{o2017oscillators}, short for swarming oscillators, which mimics various real-world systems \cite{barcis2019robots,barcis2020sandsbots,zhang2020reconfigurable}. Several researchers are now further exploring swarmalators \cite{lee2021collective,hong2018active,lizarraga2020synchronization,o2018ring,ha2021mean,sar2022swarmalators,o2019review,hong2021coupling,schilcher2021swarmalators,japon2022intercellular,vijayan2022charged}.

\begin{figure*}
    \centering
    \includegraphics[width= 2.0
    \columnwidth]{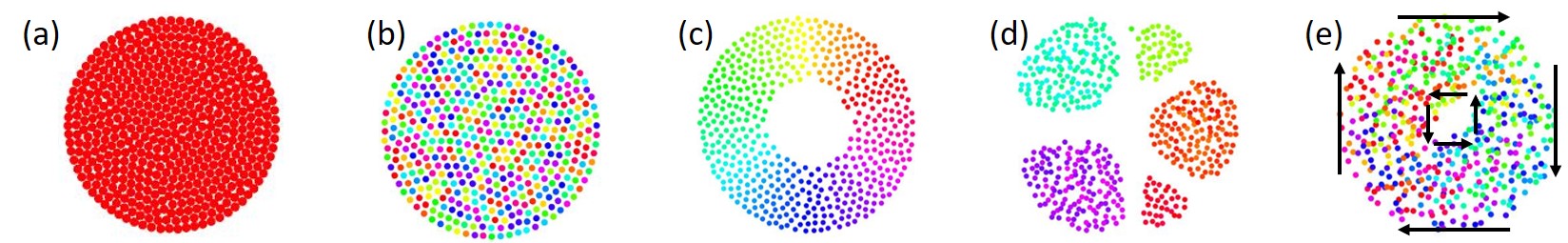}
    \caption{Collective states of the 2D swarmalator model (defined in Appendix A) where swarmalators are represented as colored dots where the color refers to the swarmalators phase. In all panels a Euler method was used with timestep $dt = 0.1 $ for $T = 1000$ units for $N = 1000$ swarmalators. (a) Static sync: $(J,K,\sigma) = (1,1,10)$ (b) Static async $(J,K,\sigma) = (1,1,10)$ (c) Static phase wave $(J,K,\sigma) = (1,1,10)$ (d) Splintered phase wave $(J,K,\sigma) = (1,1,10)$ (e) Active phase wave. In the three static states (a)-(c) swarmalators do not move in space or phase. In the splintered phase wave, each colored chunk is a vortex: the swarmalators librate in both space and phase. In the active phase wave, the librations are excited into rotations; the swarmalators split into counter-rotating groups as indicated by the black arrows. Figure adapted from \cite{o2022collective}.} 
    \label{states-2D}
\end{figure*}

\begin{figure}
    \centering
    \includegraphics[width= 0.85 \columnwidth]{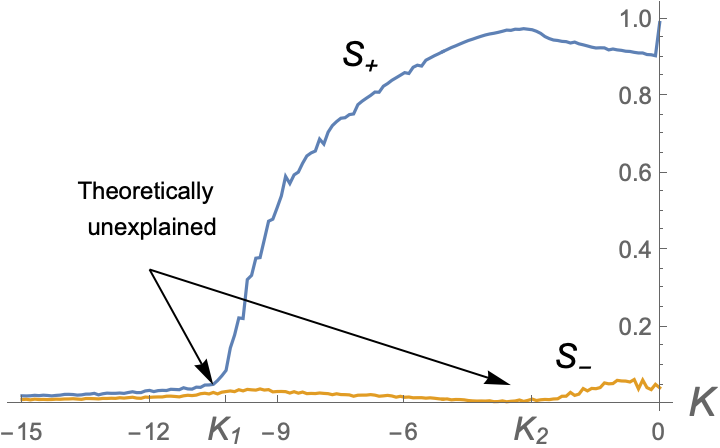}  
    \caption{Order parameters of the 2D swaramalator model (Appendix A) $S_{\pm} e^{i \phi_{\pm}} := (N)^{-1} \sum_j e^{i(\phi_j \pm \theta_j)}$, where $\phi, \theta$ are the spatial angle and phase of swarmalators.} 
    \label{order-parameters-2D}
\end{figure}

The physics of the swarmalator model \cite{o2017oscillators} is not yet understood. Varying one parameter produces five collective states (Fig.~\ref{states-2D}); the three static states (Fig.~\ref{states-2D}(a)-(c)) have been analyzed \cite{o2017oscillators}, but the two dynamical states (Fig.~\ref{states-2D}(d)-(e)) remain murky -- What is the nature of the flow in the vortex like active phase wave (Fig.\ref{states-2D}(d))? Does it imitate the flow in vortices of Janus crystals and sperm \cite{riedel2005self,yan2015rotating}? What determines the number of mini-vortices in the splintered phase wave (Fig.~\ref{states-2D}(e))? Do they mimic the rotating flocks seen in active fluids \cite{zhang2020reconfigurable}? The stabilities and bifurcations of \textit{all} states are also a mystery. Fig.~\ref{order-parameters-2D} illustrates the bifurcation structure by plotting the order parameters $S_{\pm}$ (defined later) versus the phase coupling $K$. At an unknown $K_1$, $S_+$ jumps from 0 as the async state transitions to the active phase wave. At a later $K_2$, $S_+$ begins to decline as the splintered phase wave is born. Like the old puzzles about the Kuramoto model \cite{kuramoto2003chemical,strogatz2000kuramoto,strogatz1991stability,mirollo2007spectrum,crawford1994amplitude}, the bifurcations of the swarmalator model ``cry out for [theoretical] explanation" \cite{strogatz2000kuramoto, o2022collective}.

This work is a single step in a longer journey to provide such an explanation \cite{o2022collective,yoon2022sync}. Our dream is to repurpose the tools from the sync world (Kuramoto's self-consistency analysis \cite{kuramoto2003chemical}, or perhaps even OA theory 
\cite{ott2008low}) to derive expressions for $K_1, K_2$, and hopefully some results on the stability of the static async state too.

But it's not clear how this can be done. Take finding $K_1$, the point at which async destabilizes. For the Kuramoto model, this is derived by exploiting the fact that in the sync state, the (non-identical) oscillators split into two groups, one locked at fixed points $\theta_i^*(\omega_i)$ with density $\rho_{locked}(\theta)$, the other drifting $|\dot{\theta_i}| > 0$ with $\rho_{drift}(\theta)$. Skipping over details \cite{kuramoto2003chemical}, the key to the analysis is $\rho_{drift}(\theta)$ cancels out and that $\rho_{locked}(\theta)$ has a simple form since it represents oscillators sitting at fixed points. For the swarmalator model, however, swarmalators are identical, and post-transition, \textit{non-stationary}. This implies that they have a common $\rho(x,\theta)$ (since they are identical) and that the form of $\rho(x,\theta)$ cannot be easily guessed (since they are non-stationary). So Kuramoto's trick cannot be straightforwardly adapted. 

Blocked by these mathematical walls, we took the natural back path of divide and conquer: we split the swarmalator model into its radial and angular components; if the original model in Cartesian coordinates has form $(\dot{x_i},\dot{y_i},\dot{\theta_i})$, the phase being $\theta_i$, then the radial component of the model is $(\dot{r_i}, \dot{\phi}_i)$, and the angular $(\dot{\phi}_i, \dot{\theta}_i)$, where $(r,\phi)$ are polar coordinates. The essence of the angular piece is especially simple (Appendix A),
\begin{align}
    \dot{\phi_i} &= \omega(r_i) +  \frac{1}{N} \sum_j^N J(r_i, r_j) \sin(\phi_j - \phi_i) \cos(\theta_j - \theta_i) \label{eom-x1} \\
    \dot{\theta_i} &= \nu(r_i) + \frac{1}{N}  \sum_j^N K(r_i,r_j) \sin(\theta_j - \theta_i ) \cos(\phi_j - \phi_i ) \label{eom-theta1}
\end{align}
It is a pair of Kuramoto models where now the natural frequencies and couplings depend on the $r_i, r_j$, and the familiar sine terms are modifed by cosines. The effect of the cosines is to make the sync position-dependent and the swarming phase-dependent -- a lovely symmetry which captures the raw essence of swarmalators.

This emergence of this `ring model' from the 2D model got us excited. It hinted that the tools from sync studies might indeed be adapted for these new puzzles about swarmalators. The Kuramoto model with couplings distributed as $K_i, K_j, K_{ij}$, for example, have been solved exactly \cite{hong2011conformists,hong2012mean,kloumann2014phase} -- could we adapt these works to the ring model (since it has similar form)?

This paper is the third in a series of papers which explore this tantalizing prospect. The strategy is to study the ring model piece by piece.  First we set the natural frequencies and couplings at constants $(\omega, \nu, J, K)$ \cite{o2022collective}. Then we turned on quenched disorder in $(\omega_i, \nu_i)$, keeping $(J,K)$ constant \cite{yoon2022sync}. Here we isolate $K_j$- distributed couplings (defind in model below) and keep the frequencies frozen at constants $(\omega, \nu)$. We find several new collective states, as well as generalizations of previously reported states, some of which we are able to analyze.

Lastly, we mention that the ring model with distributed $K_j$ is worth studying in its own right, and not just as a warm up for the 2D swarmlator model. It is a toy model for the  many natural swarmalators which move in quasi-1D rings such a vinegar eels and sperm \cite{bau2015worms,yuan2015hydrodynamic,ketzetzi2021activity,creppy2016symmetry,aihara2014spatio}. Asymmetric couplings, as encoded by $K_j$, are common in such systems \cite{liebchen2021interactions}, yet are rarely studied.

\section{Model}
The ring swarmalator model we study is
\begin{align}
    \dot{x_i} &= \omega +  \frac{1}{N} \sum_j^N J_j \sin(x_j - x_i) \cos(\theta_j - \theta_i) \label{eom-x} \\
    \dot{\theta_i} &= \nu + \frac{1}{N}  \sum_j^N K_j \sin(\theta_j - \theta_i ) \cos(x_j - x_i ) \label{eom-theta}
\end{align}
\noindent
where $(x_i, \theta_i) \in (S^1, S^1)$ are the position and phase of the $i$th swarmalator for $i = 1, \dots, N$ and $(\omega, \nu)$ and $(J_j,K_j)$ are the associated natural frequencies and couplings. Notice we have switched $\phi_i \rightarrow x_i$ to make it clear that $x_i$ denotes an angle in space, as opposed to an internal phase like $\theta$. We set $(\omega, \nu) = (0,0)$ via a change of frame without loss of generality (wlog). As for the $J_j, K_j$, we derive most of our results for arbitrary distributions $g(J), h(K)$, but we use a simpler `double delta` distribution
\begin{align}
g(J) &= \delta(J-1) \\ 
h(K) &= p\delta(K-K_p)+(1-p)\delta(K-K_n)
\label{double_delta}
\end{align}
where $K_p > 0$ and $K_n < 0$ and $0 \leq p \leq 1$, as a working example throughout.

\section{Numerics}
We use two order parameters to catalog our models macroscopic behavior:
\begin{align}
  & W_{\pm} = S_{\pm}e^{i\Phi_{\pm}} \equiv \frac{1}{N}\sum_{j=1}^N e^{i(x_j\pm \theta_i)} \\
  &  V  = \frac{1}{N} \sum_{j=1}^N \Big\langle\sqrt{{\dot{x_j}}^2+{\dot{\theta_j}}^2}\Big\rangle_t,
\end{align}
The $S_{\pm}$ `rainbow order parameters` -- so-called since they are maximal in the rainbow like static phase wave state; Fig.~\ref{states-2D}(c) -- measure the global space-phase order. They are maximal $S_{\pm} = 1$ when $x_i = \pm \theta_i + C$ for some constant $C$. They are minimal $S_{\pm} = 0$ when $x_i$ and $\theta_i$ are uncorrelated. These order parameters arise naturally in the ring swarmalator model \cite{o2017oscillators} and can distinguish between most of the model's emergent states. They are blind, however, to swarmalators motion. So we use the mean velocity $V$ to detect if a collective state exists in which swarmalators are moving.

\begin{figure}
    \centering
    \includegraphics[width= \columnwidth]{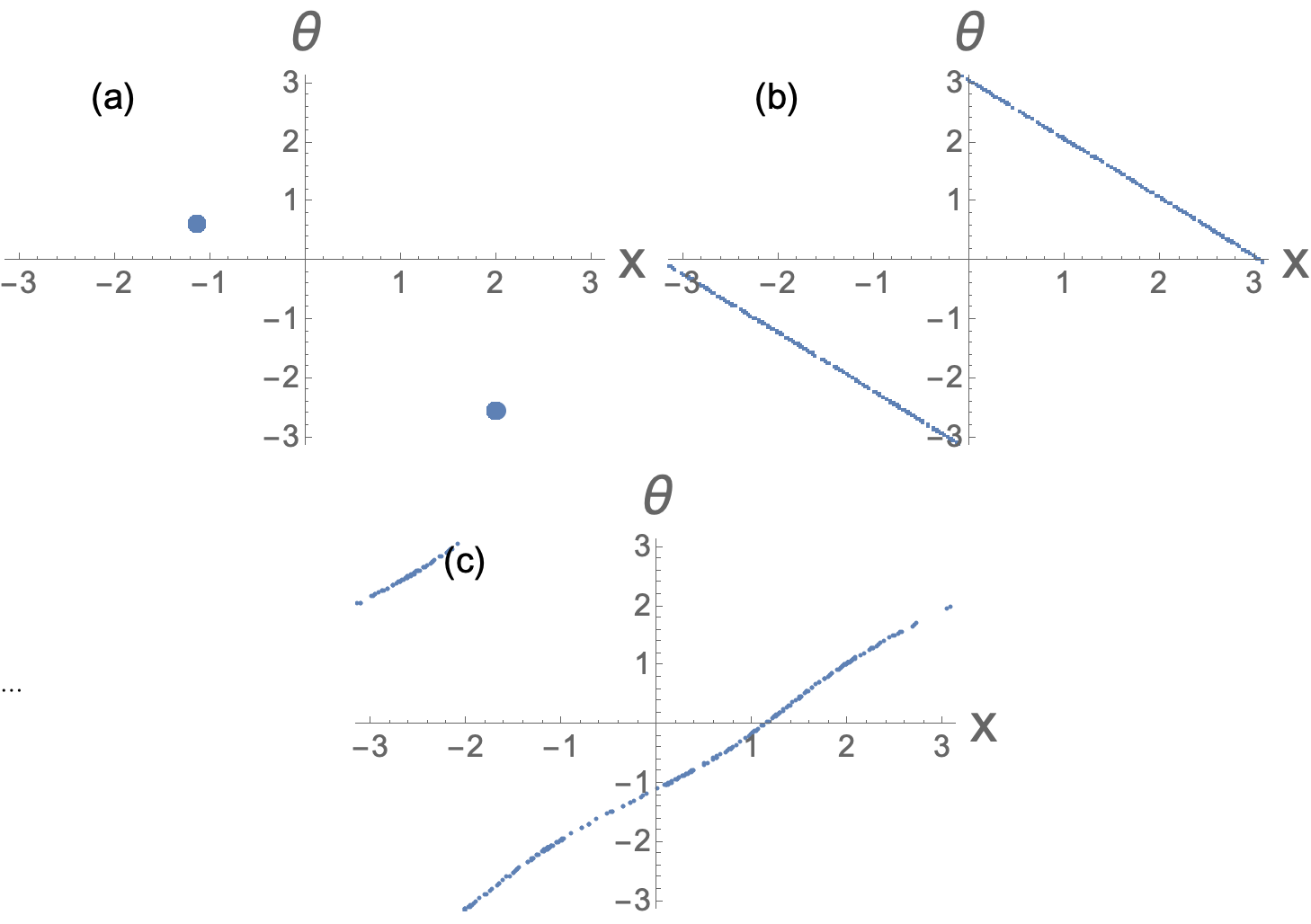}  
    \caption{Static collective states as scatter plots in $(x,\theta)$ plane. (a) Static sync, $(p,K_p, K_n) = (0.3, -0.5, 2)$; (b) Static phase wave, $(p,K_p, K_n) = (0.9, -0.5, 2)$; (c) Buckled phase wave, $(p,K_p, K_n) = (0.5, 1, -1.5)$;  Simulation parameters: RK4 method with $(dt, T, N) = (0.1, 100, 500)$. In (c), $N = 300$. In (a), the point sizes have been enlarged to make things clearer.} 
    \label{static_states}
\end{figure}

\begin{figure}
    \centering
    \includegraphics[width= \columnwidth]{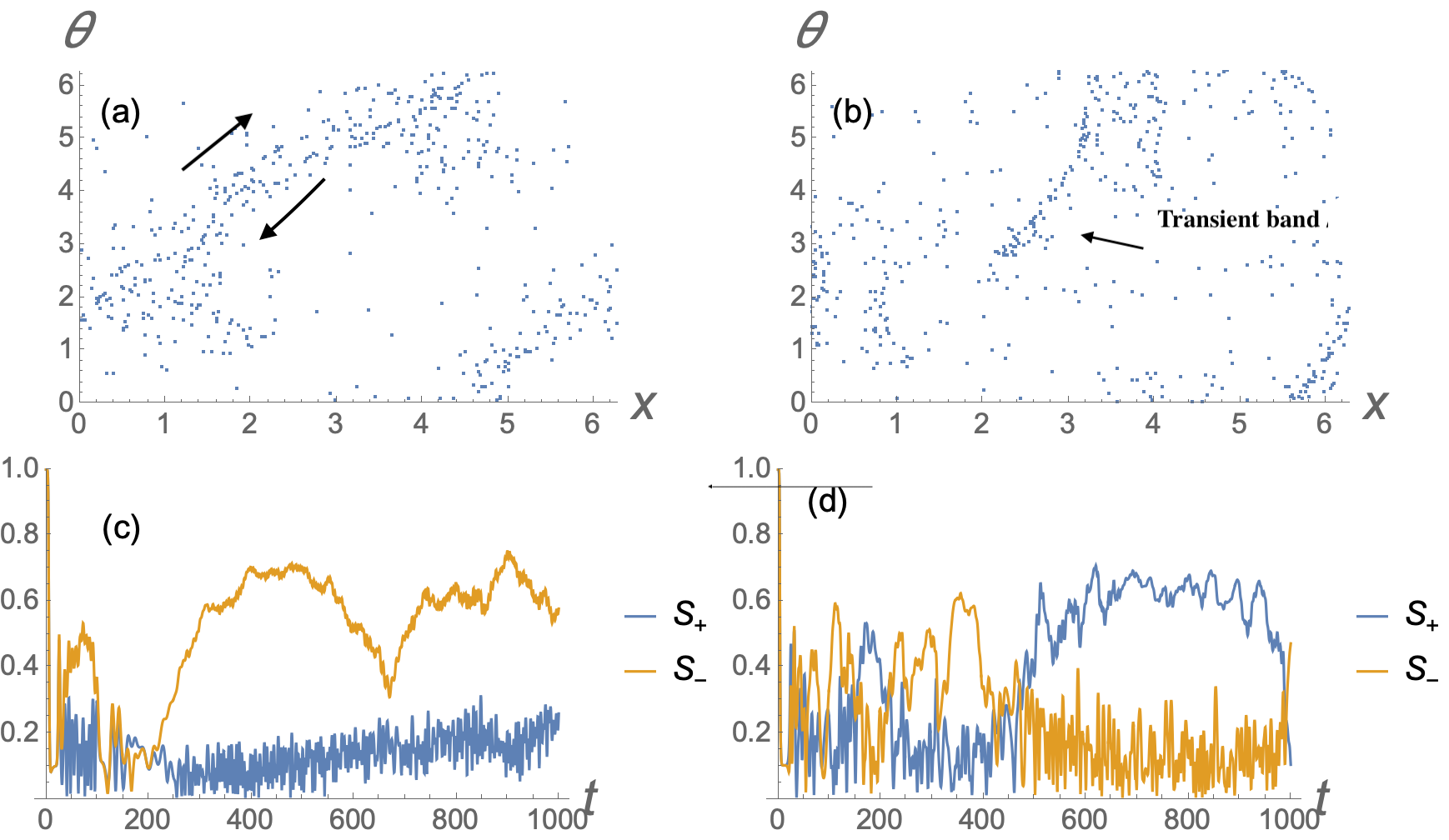}
    \caption{Unsteady collective states. Top row, scatter plots in $(x,\theta)$ space. Bottom row, time series of order parameters. (a)-(c) Swarmalators execute shear flow as indicated by the black arrows, like in the active phase wave on the 2D swarmalators model, but now $S_{\pm}$ have noisey oscillations. Here $(p,K_1,K_2) = (0.3,1,-2)$ (b)-(d). Swarmalators execute erratic gas like motion with $S_{\pm}$ both noisey and similar magnitude. Bands of ordered swarmalators spotaneously appear then disappear. Here $(p, K_1, K_2) = (0.45, 5, -7)$ (best viewed in Supplementary Movie 1). Sim parameters: $(dt, T, N) = (0.25, 2000, 500)$.} 
    \label{dirty_phase_wave}
\end{figure}

We numerically integrated the governing equations using RK4 method and found five collective states. Figure~\ref{S_Q2} plots $S(p),V(p)$ which demarcates the states. Code used for simulations is available at \cite{repo},  movies of all states available in SM. The states are:

\begin{itemize}
\item \textbf{Static sync} for $p>p_c$: $(x_i, \theta_i) = (x^*, \theta^*)$. A $`\pi$-state` where $(x_i, \theta_i) = (x^*, \theta^*) \cup (x^* + \pi, \theta^* + \pi)$ is found too \cite{o2022collective}. We call them both static sync. Order parameters are $S_{\pm}=1$ and $V=0$ (Fig~\ref{static_states}(a)).
\item \textbf{Static phase wave} for $p_0<p<p_c$: $x_i = \theta_i \pm C$ (the $\pm$ refers to a clockwise or counter-clockwise phase wave). Here either $(S_+, S_-) = (1,0)$ or $(0,1)$ (depending on the $\pm$) and $V=0$. Realized as $N \rightarrow \infty$ (Fig~\ref{static_states}(b)).
\item \textbf{Buckled phase wave} near $p_c$: A static phase wave with a 'buckle' so $S_+ \approx 1, S_- = 0, V = 0$. Realized for finite $N$ (Fig~\ref{static_states}(c)).
\item \textbf{Noisy phase wave}  for $p<p_0$: the static phase wave destabilizes into noisy, unsteady phase waves with $V >0$. For $p$ near $p_c$, there is approximate shear flow (Fig~\ref{dirty_phase_wave}(a)) similar to the active phase wave of the 2D model (Fig~\ref{states-2D}(c)). Here however, the space correlation between $x_i \approx \theta_i$ fluctuates as illustrated by the noisy $S_{\pm}$ time series (Fig~\ref{dirty_phase_wave}(c)) where  $S_+ > 1, S_- \approx 0$. Realised for finite $N$.
\item \textbf{Async} for $p \approx 0$: For smaller $p$, the shear flow degenerates into erratic gas like motion (Fig~\ref{dirty_phase_wave}(b)) with both $S_+, S_-$ noisy (Fig~\ref{dirty_phase_wave}(d)) which we call `active async'. Bands of sync'd swarmalators spontaneously appear and disappear (best viewed in Supplementary Movie 1). As $p \rightarrow 0$, $S_{\pm}, V \rightarrow 0$ gradually decline indicating the system becomes fully incoherent. Strangely, for all finite $N$ we probed (up to $10^4$ swarmalators) the async state is `active` with small but finite mean velocity $V > 0$. In the continuum limit $N \rightarrow \infty$, however, the state becomes truly static $V = 0$ (we prove this later).
\end{itemize}
\section{Analysis}
\subsection{Static sync}
Here swarmalators sit at fixed points: $(x_i, \theta_i) = (x^*,\theta^*)$. The $\pi$ state, in which swarmalators split into two groups, one at $(x^*, \theta^*)$, the other at $(x^* + \pi, \theta^* + \pi)$, is dynamically equivalent to the single cluster state because the governing ODEs are invariant under the `$\pi$-transformation' $x,\theta \rightarrow x + \pi, x + \pi$ \cite{yoon2022sync}. So we analyze the one cluster state in which $(x_i, \theta_i) = (x^*, \theta^*)$ without loss of generality (wlog).

Now we derive the stability of this state for arbitrary $g(J), h(K)$. Linearizing Eqs.~(\ref{eom-x}) and (\ref{eom-theta}) about this fixed point yields
\begin{equation}
    \left[ 
\begin{array}{c} 
  \dot{x_i} \\ 
  \dot{\theta_i}\\
\end{array} 
\right] 
=M
\left[ 
\begin{array}{c} 
  x_i \\ 
  \theta_i\\
\end{array} 
\right], \label{eom-matrix}
\end{equation}
where 
the Jacobian $M$ for the static sync at this fixed point has a block structure:
\begin{figure}
    \centering
   \includegraphics[width= 0.95 \columnwidth]{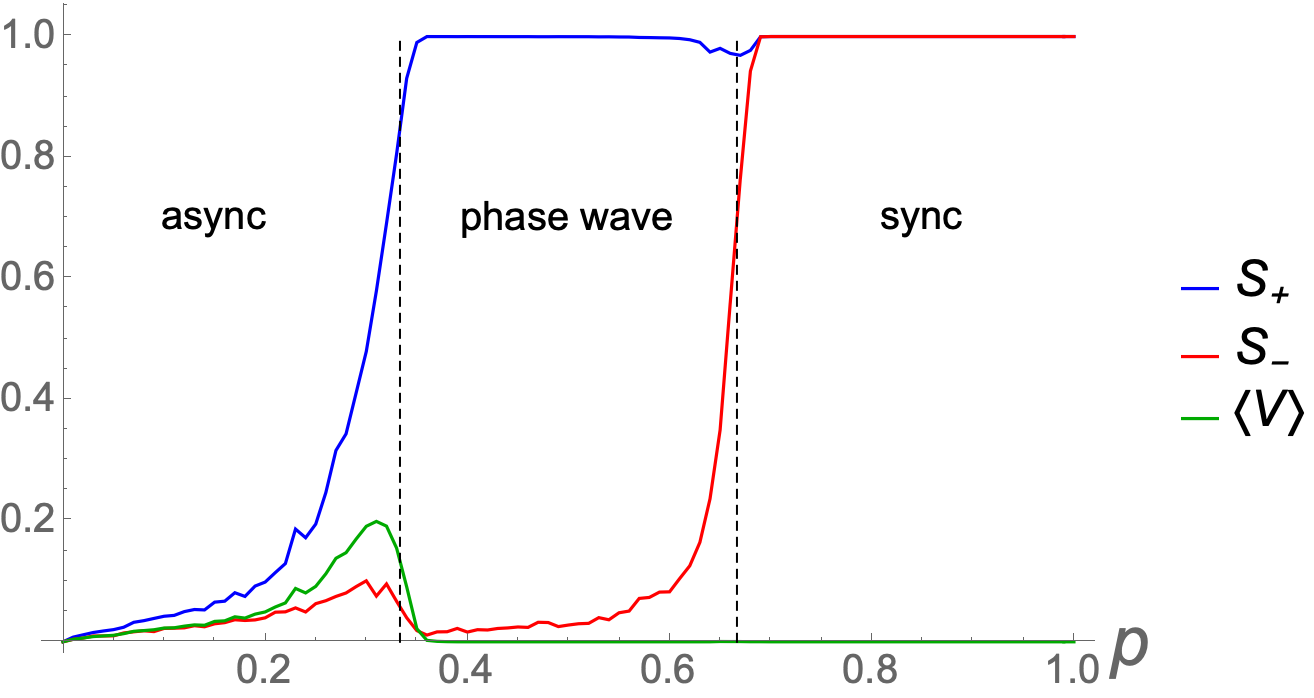}
   \caption{Order parameters $S_{\pm}$ and $V$ as a function of $p$ for $Q=2$ with $Q = -K_n / K_p$. Critical $p_c, p_s$ are given by Eq.~\eqref{p_c}, Eq~\eqref{p_s}. For $p_s \leq p \leq p_c$, the buckled phase wave is realized for finite $N$ for which $S_- >0$. As $N \rightarrow \infty$, the buckle disappears and the static phase wave is realized in which $S_{-} = 0$. We assume $S_{+} > S_-$ wlog which amounts to studying the clockwise, as opposed to counter-clockwise, phase wave (see text). For $p < p_s$ and finite $N$, the active async state is realized with $V > 0$. As $N \rightarrow \infty$, $V \rightarrow \infty$ and the static async state is born. Simulation parameters: RK4 method with $(dt, T, N) = (0.01, 1000, 1600)$. Each data point is the average of 20 samples.} 
    \label{S_Q2}
\end{figure}
\begin{equation}
    M = \frac{1}{N}\left[ 
\begin{array}{cc} 
  A & 0 \\ 
  0 & B \\
\end{array} 
\right], \label{Mss}
\end{equation}
where
\begin{equation}
A := \begin{bmatrix}
- \sum_{j\neq 1} J_j  & J_2  & \dots & J_N \\
    J_1  & - \sum_{j\neq 2} J_j &  \dots & J_N \\
    \vdots & \vdots &  & \vdots \\
    J_1  & J_2   & \dots & - \sum_{j \neq N} J_j 
\end{bmatrix} \label{A} 
\end{equation}
and
\begin{equation}
B := \begin{bmatrix}
- \sum_{j\neq 1} K_j  & K_2  & \dots & K_N \\
   K_1  & - \sum_{j\neq 2} K_j &  \dots & K_N \\
   \vdots & \vdots &  &\vdots \\
   K_1  & K_2  & \dots & -\sum_{j \neq N} K_j 
\end{bmatrix}. \label{B}
\end{equation}
The matrices $A$ and $B$ have been studied before \cite{o2017oscillators}. Their eigenvalues are $\lambda_A = 0, -\langle J \rangle$ and $\lambda_B = 0, - \langle K \rangle$ with multiplicities $1, N-1$ (the zero eigenvalues stem from the rotational symmetry in the model). The eigenvalues of $M$ are the union of those of $A$ and $B$  $ \lambda_M = \lambda_A \cup \lambda_B$. This follows from $M$'s block structure: $det(M) = det(A) det(B)$. Putting this together gives
\begin{align}
    \lambda_0 &= 0 \hspace{1 cm} (\text{w.m.} \; \; 2) \\
    \lambda_1 &= -\langle J \rangle  \hspace{1 cm} (\text{w.m.} \; \; N-1) \\
    \lambda_2 &= -\langle K \rangle \hspace{1 cm} (\text{w.m.} \; \; N-1) 
\end{align}
where $w.m.$ denotes \underline{w}ith \underline{m}ultiplicity. Thus sync destabilizes when $\langle K \rangle = 0 $ or $\langle J \rangle = 0$  which, recall, holds for \textit{general} $g(J), h(K)$. For the double delta distribution working example  (Eq.~\eqref{double_delta}),
\begin{align}
    \langle K \rangle &= p K_p + (1-p) K_n \\
    &= K_p [ p(1+Q) - Q ] \nonumber,
\end{align}
where $Q\equiv -K_n/K_p$.  Setting this to zero gives the critical fraction of positively coupled swarmlators
\begin{equation}
    p_c = \frac{Q}{1+Q}.
    \label{p_c}
\end{equation}

\subsection{Buckled phase wave}
For $p < p_c$ and finite $N$ the buckled phase wave is born. Here we derive the the 1D manifold $\Gamma(x, \theta) = 0$ which defines the state for arbitrary $g(J), h(K)$ (the stability of the state is out of scope).

First we move to $(\xi, \eta)$ coordinates
\begin{align}
    \xi_i = x_i + \theta_i \\
    \eta_i = x_i - \theta_i 
\end{align}
The governing equations become
\begin{align}
    \dot{\xi_i} &= \frac{U_+}{2} \sin( \Psi_+ - \xi_i) + \frac{V_+}{2} \sin( \Phi_+ - \xi_i) \nonumber  \\
                & + \frac{U_-}{2} \sin( \Psi_+ - \eta_i) - \frac{V_-}{2} \sin( \Phi_- - \eta_i) \label{eom_xi} \\
    \dot{\eta_i} &= \frac{U_+}{2} \sin( \Psi_+ - \xi_i) - \frac{V_+}{2} \sin( \Phi_+ - \xi_i) \nonumber \\
                & + \frac{U_-}{2} \sin( \Psi_+ - \eta_i) + \frac{V_-}{2} \sin( \Phi_- - \eta_i) \label{eom_eta}
\end{align}
where
\begin{align}
    U_{\pm} e^{i \Psi_{\pm}} = \frac{1}{N}\sum_j J_j e^{i (x_j \pm \theta_j)}, \\
    V_{\pm} e^{i \Phi_{\pm}} = \frac{1}{N}\sum_j K_j e^{i (x_j \pm \theta_j)}. 
\end{align}
are `glassy' order parameters \cite{kloumann2014phase}. Next set Eq.~\eqref{eom_xi},\eqref{eom_eta} to zero since swarmalators are at fixed points. Then we add and subtract the equations to produce
\begin{align}
    0 &= U_+ \sin \xi_i + U_- \sin \eta_i,  \label{q3} \\
    0 &= V_+ \sin( \Phi_+ - \xi_i) - V_{-} \sin( \Phi_- - \eta_i). \label{q4}
\end{align}
where we set $\Psi_{\pm} = 0$ wlog. Eqs.~\eqref{q3},~\eqref{q4} are nullclines, curves in $(\xi, \eta)$ space, $\Gamma_1(\xi, \eta) = 0, \Gamma_2(\xi,\eta) = 0$. Observe that (i) The nullclines must be identical and (ii) describe the buckled phase wave $\Gamma_1 = \Gamma_2 = \Gamma$. (i) implies
\begin{align}
\frac{U_-}{U_+} = \frac{V_-}{V_+} \\
\Phi_{+} - \Phi_{-} = \pi
\end{align}
which we have confirmed numerically. (ii) implies 
\begin{equation}
    \Gamma(\xi,\eta) = \sin \xi + u \sin \eta = 0 \label{nullcline}
\end{equation}
where $u := U_- / U_+$ and we have abused notation by using $\Gamma$ for the curve in $(\xi,\eta)$ space: $\Gamma(\xi, \eta) \Longleftrightarrow \Gamma(x,t)$. 

\begin{figure}
    \centering
    \includegraphics[width= \columnwidth]{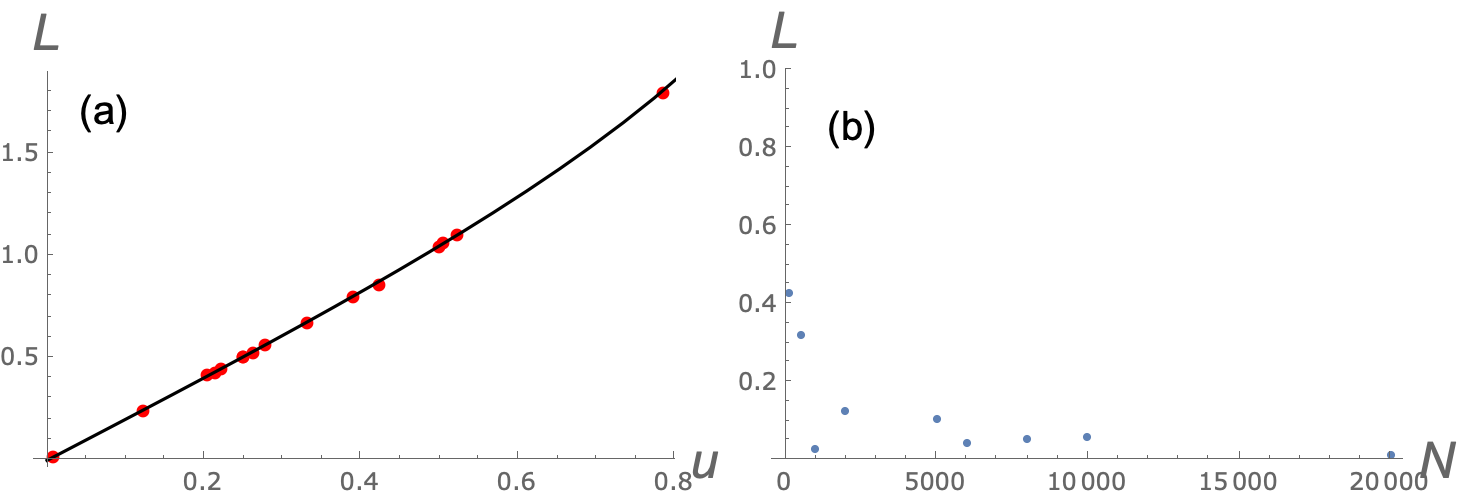}
    \caption{(a) Black line, prediction Eq.~\eqref{width}. Red dots, simulation parameters for $(p,K_p,K_n) = (0.5,1,-1.5)$. Sim pars $(dt,T) = (0.25,1000)$ for $N = 5, 10, \dots 100$. Any simulations for which the buckled phase wave was not realized was discarded. We asserted $S_+ > S_-$ in simulations wlog for that $U_+ > U_- \rightarrow u < 1$. (b) Buckle size approaches $0$ for large $N$.} 
    \label{L-u}
\end{figure}

This is the desired parameterization of the buckled phase wave in terms of the glassy order parameters, $U_{\pm}$. We tested Eq.~\eqref{nullcline} as follows. Let the buckled be in the $\xi$ direction wlog and define its size $L := \max(\xi_i) - \min(\xi_i)$. The buckle is symmetric about $\xi = 0$ (really about $\Psi_+ = \langle \xi \rangle$ which we set to $0$) so $(\xi,\eta) = (\max(\xi_i) = L/2,\pi/2)$ lies on $\Gamma$. Then Eq.\eqref{nullcline} implies
\begin{equation}
    L = 2 \arcsin{u} \label{width}
\end{equation}
We confirmed this prediction by simulating the system for various $N$ and numerically computing $u$ as depicted in Figure~\ref{L-u}(a).

We can compute $u$ analytically as $N \rightarrow \infty$ using Kuramoto's self-consistency trick \cite{kuramoto2003chemical}. Figure~\ref{L-u}(b), however, shows the buckle disappears as $N \rightarrow \infty$ (the static phase wave is approached) which implies $(U_+, U_-) = (1,0) \Rightarrow u = 0$; so the calculation is in a sense moot. Nevertheless, we include it to show how the self-consistency calculation works for a density $\rho(\xi,\eta)$ that is defined on compactly supported, non-trivial manifold $\Gamma(\xi,\eta) = 0$ (as opposed to being defined on a fully supported space, like the density of oscillators $\rho(\theta)$ of the Kuramoto model which lives on $S^1$). 

As $N \rightarrow \infty$, the expressions for the glassy order parameters become
\begin{align}
    & U_+ = \int_{\Gamma(\xi,\eta)} J \cos \xi \rho(\xi, \eta, J, K) g(J) h(k) d \xi d \eta dJ dK \label{t1} \\
    & U_- = \int_{\Gamma(\xi,\eta)} J \cos \eta \rho(\xi, \eta, J, K) g(J) h(k) d \xi d \eta dJ dK \label{t2} \\
    & \Gamma(\xi,\eta, u) := \sin \xi + u \sin \eta = 0 \label{t3} 
\end{align}
where we have set $\Psi_{\pm} = 0$ wlog (which means the integrands $e^{i \xi}, e^{i \eta} \rightarrow \cos \xi, \cos \eta$) and $\rho(\xi, \eta, J, K)$ is the density of swarmalators in the Eulerian sense. Eq.~\eqref{t3} requotes the definition of $\Gamma(\xi,\eta,u)$ for convenience, and explicitly denote its dependence on $u$.

Notice in contrast to the self-consistency equations for the regular Kuramoto order parameter \cite{kuramoto2003chemical}, the integrals above are \textit{contour} integrals over $\Gamma(x,\theta,u)$. And crucially, $\Gamma(\xi, \eta, u)$ depends on $u$. So Eqs.~\eqref{t1}-\eqref{t3} are a set of self-consistency equations for \textit{four} quantities: $U_+, U_-, \rho(\xi, \eta, J, K), \Gamma(\xi, \eta, u)$ -- quite a challenge!

Let's break them down. Recall the contour integral of a function $f(x,y)$ over a curve $\delta(s)$ is
\begin{align}
I &= \int_{\delta} f(x,y) ds  \\
I &= \int_{s_1}^{s_2} g(s) \sqrt{1 + \delta'(s)^2} ds 
\end{align}
where $g(s) := f(x(s), y(s))$ is the function evaluated along the contour $\delta(s)$ which has extremal points $(s_1, s_2)$. Now we apply this definition to Eqs.~\eqref{t1}-\eqref{t2}. First we need an expression for the contour: $\Gamma(\xi,\eta,u) \Longleftrightarrow  \gamma(s) := \arcsin{ u \sin(s)} $ where we have chosen $s := \eta$ as the active parameter which runs from $[-\pi, \pi]$. The line measure is $\sqrt{1 + \gamma'(s)^2} = \sqrt{1+\frac{u^2 \cos ^2(s)}{1-u^2 \sin ^2(s)}} $. Then Eqs.~\eqref{t1}-\eqref{t2} become
\begin{align}
    U_+ &= \int_{-\pi}^{\pi} J \cos \xi(s) \sqrt{1 + \frac{u^2 \cos ^2(s)}{1-u^2 \sin ^2(s)}} \hat{\rho}(s,J,K) ds d\hat{J} d\hat{K} \label{r1}  \\
    U_- &= \int_{-\pi}^{\pi} J \cos s \sqrt{1 + \frac{u^2 \cos ^2(s)}{1-u^2 \sin ^2(s)}} \hat{\rho}(s,J,K) ds d\hat{J} d\hat{K} \label{r2}
\end{align}
where $\cos \xi(s) = \sqrt{1 - u^2 \sin s^2}$ is found from the definition of the contour $\sin\xi + u \sin \eta$, $\hat{\rho}(s) := \rho(\xi(s), \eta(s))$ is the (unknown) density along the contour $\gamma(s)$, and we have defined $(d\hat{J}, d\hat{K}) := (g(J) dJ, h(K) dK)$ for clarity. Recall $u := U_- / U_+$, so Eqs.~\eqref{r1},\eqref{r2} are a pair of self-consistency equations for $(U_-, U_+, \hat{\rho})$. Recall also that these are valid for arbitrary $g(J), h(K)$.

In principle, the next step is to derive an expression for $\hat{\rho}$ in terms of $U_{\pm}$ by solving the continuity equation. This is a daunting task, beyond the scope of the current paper (it's hard to do for the regular PDEs encountered in fluid mechanics, never mind the integro-PDE we are dealing with; we have an integro-PDE piece because the mean field coupling imposes non-locality in the velocity $v$). In practice, we guess an ansatz for $\hat{\rho}$. 

Numerics indicate such an ansatz is a uniform density $\hat{\rho}(s,J,K)= C^{-1}$ where $C$ is a normalization constant \footnote{Note the dependence on $(J,K)$ drops ou. There are two ways to interpret this. First, is assume that at every $s$ there is a full distribution of swarmalators with $g(J), h(K)$. In other words, the mass of $\rho$ is spread out evenly over its arguments $(s,J,K)$ support $S^1 \times \mathbb{R}$. Second, we can interpret $\hat{\rho}(s,J,K)$ as the average over manly distributions of $(J,K)$}. Let's find $U_-$. Observe that the line measure $\frac{u^2 \cos ^2(s)}{1-u^2 \sin ^2(s)}$ in the integrand is symmetric and positive definite about $0$, so when integrated against $\cos(s)$, as in Eq.~\eqref{r2}, we get $U_- = 0$. This then trivializes the calculation for $U_+$. If $U_- = 0$, then $u \rightarrow 0$ (assuming $U_+ > 0$ so denominator is not zero) and Eq.~\eqref{r1} reduces to $U_{+} = 1$. Thus,
\begin{align}
    U_+ &= 1 \\
    U_- &= 0
\end{align}
To recap, we have proved that if the density along the contour $\Gamma(\xi,\eta)$ is uniform $\hat{\rho} = C$, then $(U_-, U_+) = (0,1)$, which means the contour is a straight line $\Gamma(\xi,\eta) = \xi - \Psi_+ = 0 $ where we have reinserted $\Psi_{+}$ for clarity (remember we set $\Psi_{+} = 0$ wlog just under Eq.~\eqref{q4} \footnote{$\Phi_{+}$ is determined by the initial condition and can be to zero wlog because of the rotational symmetry of the model}). In other words, we are in the static phase wave, as we anticipated at the start of the calculation.

\subsection{Static phase}
Numerics suggests this state is unstable for all finite $N$. We were however unable to prove this. For the finite case, disordered $K_j$ made finding the eigenvalues of the associated Jacobian too difficult (for constant couplings, the eigenvalues were findable for all finite $N$! \cite{o2022collective}). For the infinite case, the density of the state is a delta function $\rho(\xi,\eta) = (2\pi)^{-1} \delta(\xi - C)$ or $(2\pi)^{-1} \delta(\eta - C)$ which is difficult to perturb off of. So instead, we numerically computed the eigenvalues for various $N$ which confirmed the state was unstable. We hope future researchers will provide a rigorous proof.

We do however have a trivial result for the glassy order parameters. Assuming the clockwise phase wave $(S_+, S_-) = (1,0)$ and the double delta coupling distribution, we get
\begin{align}
    V_+ &= \langle K e^{i \xi} \rangle = \langle K \rangle \\
    V_+ &= p(K_1 - K_2) + K_2
\end{align}
which agrees with simulation. The other order parameters are trivial:  $V_- = U_- = 0$ and $U_+ = 1$.

\subsection{Static async}
In the $N \rightarrow \infty$ limit this state is given by $\rho(\xi,\eta, K, t) = (4 \pi)^{-2}$. The density obeys the continuity equation
\begin{align}
    \dot{\rho} + \nabla . ( v \rho ) = 0 \\
    \dot{\rho} + v. \nabla \rho + \rho \nabla . v = 0 
    \label{cont1}
\end{align}
where the velocity $v$ is given by Eqns \eqref{eom_xi}, \eqref{eom_eta} and is interpreted in the Eulerian sense. Consider the perturbation
\begin{equation}
    \rho = \rho_0 + \epsilon \rho_1 = (4 \pi^2)^{-1} + \epsilon \rho_1(\xi, \eta, t) 
    \label{peturb1}
\end{equation}
We sub this perturbation into the continuity equation, expand $\rho_1$ in a Fourier Series,
\begin{align}
    \rho_1(\xi, \eta, J_+, J_-, t) &= \frac{1}{4 \pi^2} \Big( c_+(J_+, J_-, t) e^{-i \xi} + \nonumber \\ & c_-(J_+, J_-, t)e^{-i \eta} + c.c. \nonumber \\
    \rho_1^{\dagger}(\xi, \eta, J_+, J_-, t \Big) \label{rho1_fs}
\end{align}
where $\rho_1^{\dagger}$ contains all the higher harmonics, and collect the ODEs for each mode. The result is
\begin{align}
    \dot{c}_{\pm}= \frac{1}{2} \int \hat{J}_+ c_{\pm}( \hat{J}_+, \hat{J}_-, t) h(\hat{J}_+) h(\hat{J}_-) d \hat{J}_+ d \hat{J}_-
\end{align}
Seeking the discrete spectrum $c = e^{\lambda t} b(\hat{J}_+, \hat{J}_-)$ we eventually find
\begin{align}
    \lambda = \frac{\langle J^+ \rangle}{2}
\end{align}
Setting $\lambda = 0$ then yields
\begin{align}
    & \langle J^+ \rangle = 0 \\
    & \langle J \rangle + \langle K \rangle = 0 \\
    & \langle K \rangle_c = - \langle J \rangle  
\end{align}
which generalizes the finding in \cite{o2021collective}. This is for general distributions $f(J), h(K)$. For the double delta working example this becomes
\begin{equation}
p_s = \frac{1 + K_n}{K_n - K_p}
\label{p_s}
\end{equation}

\begin{figure}
   \centering
    \includegraphics[width= 0.9 \columnwidth]{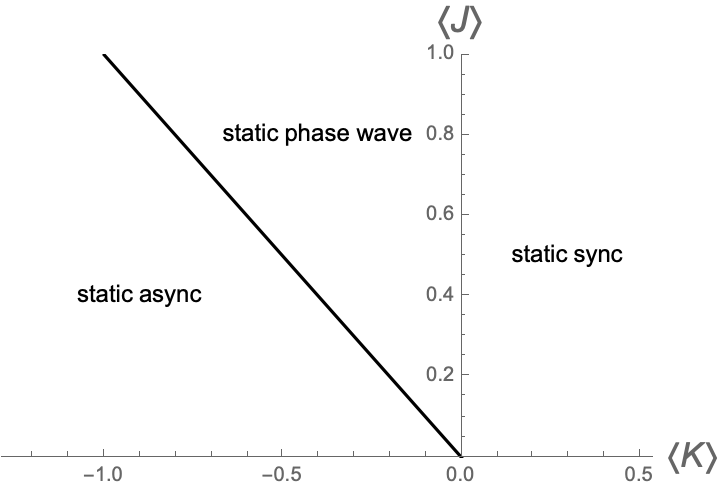}
    \caption{Bifurcation diagram in $(\langle J \rangle, \langle K \rangle)$ space in the $N \rightarrow \infty$ limit valid for arbitrary coupling distributions $g(J), h(K)$.} 
    \label{bif_diagram}
\end{figure}

This completes our analysis. Figure~\ref{bif_diagram} reports the bifurcation diagram in $(\langle J \rangle, \langle K \rangle)$ space which is valid in the $N \rightarrow \infty$ and for \textit{arbitrary} coupling distributions $g(K), h(K)$. It is a clean generalization of the identical coupling limit in \cite{o2022collective} where the couplings are replaced by their mean values $J,K \rightarrow \langle J \rangle, \langle K \rangle$. 

\section{Discussion}
The ring model is intended as a stepping stone to the 2D swarmalator model. Previous work showed constant couplings produced 1D analogues of the static sync, static phase wave, and static async states \cite{o2022collective}, while distributed $(\nu_i, \omega_i)$ produced a 1D active phase wave \cite{yoon2022sync}. We were hoping $K_j$-couplings might produce a 1D splintered phase wave (specifically: a statistically stationary, and thus analyzable, 1D analogue.)

This wasn't the case. For infinite $N$, we found the same physics as contant coupling model: the self-same static states cropped up. Moreover, the critical couplings were simply promoted to averages; the sync boundary became $(J>0, K>0) \rightarrow (\langle J \rangle > 0, \langle K \rangle > 0)$, the static async $J_+ <0 \rightarrow \langle J_+ \rangle  < 0$. Still, this negative result is useful. It tells us that $K_j$ is not the mechanism behind the non-stationarity in the splintered phase wave. Our next work will investigate if  $K_i$-type couplings (where the $K_i$ sits outside the sum in Eq.~\eqref{eom-x},\eqref{eom-theta}) will produce a 1D splintered phase wave.

Recall for finite $N$, however, the $K_j$-couplings \textit{did} produce new physics: a noisy phase wave and an active async state. The shear flow in the noisy phase wave imitates the flow in real world swarmalators such as sperm \cite{creppy2016symmetry} and vinegar eels \cite{quillen2021metachronal,quillen2021synchronized} which spontaneously form counter-rotating phases waves when confined to quasi 1D ring-like geometries. The interesting part (to us at least) is that the noisy behavior occurs for \textit{identical} swarmlators; the fluctuations arise from neither heterogeneous natural frequencies nor from external sources, as one might expect. Rather, they are generated from the interactions between the internal and external degrees of freedom (phases $\theta_i$ and positions $x_i$ respectively). As for the active async state, the transient cluster formation was not seen in the constant coupling model, and is reminiscent of the cluster dynamics in recent experiments of synthetic microswimmers \cite{ketzetzi2022activity} (these cluster dynamics are best viewed in Supplementary Movie 1). 

Opportunities for future work include adding delayed interactions, external forcing, or heterogeneous natural frequencies. A proof for the stability of the static phase wave (see Section IV.C) would also be interesting.


\section{Acknowledgements}
This research was supported by the NRF Grant No. 2021R1A2B5B01001951 (H.H).

\newpage


\appendix

\newpage

\begin{widetext}
\section{Connection of ring model to 2D swarmalator model}

Here we show how the ring model is contained within the 2D swarmalator model which is given by
\begin{align}
&\dot{\mathbf{x}}_i = \mathbf{v}_i + \frac{1}{N} \sum_{j=1}^N \Big[ \mathbf{I}_{\mathrm{att}}(\mathbf{x}_j - \mathbf{x}_i)F(\theta_j - \theta_i)  - \mathbf{I}_{\mathrm{rep}} (\mathbf{x}_j - \mathbf{x}_i) \Big],   \\ 
& \dot{\theta_i} = \omega_i +  \frac{K}{N} \sum_{j=1}^N H_{\mathrm{\mathrm{att}}}(\theta_j - \theta_i) G_{\sigma}(\mathbf{x}_j - \mathbf{x}_i)
\end{align}
In \cite{o2017oscillators}, the choices $I_{att} = x / |x|$, $I_{rep} = x / |x|^2$, $F(\theta) = 1 + J \cos(\theta)$, $G(x) = 1 / |x|$, $H_{att}(\theta) = \sin(\theta)$ were made. However, choosing linear spatial attraction $I_{att}(x) = x$, inverse square spatial repulsion $I_{rep}(x) = x / |x|^2$ and truncated parabolic space-phase coupling $G(x) = (1 - |x|^2/ \sigma^2) H_{heaviside}({\sigma - |x|})$ 
\begin{align}
&\dot{\mathbf{x}}_i = \frac{1}{N} \sum_{ j \neq i}^N \Bigg[ \mathbf{x}_j - \mathbf{x}_i \Big( 1 + J \cos(\theta_j - \theta_i)  \Big) -   \frac{\mathbf{x}_j - \mathbf{x}_i}{ | \mathbf{x}_j - \mathbf{x}_i|^2}\Bigg] \label{linear_parabolic1} \\ 
& \dot{\theta_i} = \frac{K}{N} \sum_{j \neq i}^N \sin(\theta_j - \theta_i ) \Big(1 - \frac{| \mathbf{x}_j - \mathbf{x}_i|^2}{\sigma^2} \Big) H_{heaviside}(\sigma - |\mathbf{x}_j - \mathbf{x}_i|) \label{linear_parabolic2}
\end{align}
\noindent
gives the same qualitative behavior but is nicer to work with analytically (see Appendix in \cite{o2022collective})

The `linear parabolic` model, so called because $I_{att} = x$ and $G(x)$ is a parabolic, is cleaner analytically. In polar coordinates it takes form

\begin{align*}
\dot{r_i}&=  H_r(r_i, \phi_i) - J r_i R_0 \cos \Big( \Psi_0 - \theta_i \Big) + \frac{J}{2} \Bigg[ \tilde{S}_+ \cos \Big( \Phi_+- (\phi_i+\theta_i) \Big) 
 +  \tilde{S}_{-}\cos \Big( \Phi_{-} - (\phi_i-\theta_i) \Big) \Bigg] \\
\dot{\phi_i}&=  H_{\phi}(r_i, \phi_i)  + \frac{J}{2 r_i} \Bigg[ \tilde{S}_+ \sin \Big( \Psi_+ - (\phi_i+\theta_i) \Big)  + \tilde{S}_- \sin \Big( \Psi_{-} -  (\phi_i- \theta_i) \Big) \Bigg] \\
\dot{\theta_i} &=  K \Big(1- \frac{r_i^2}{\sigma^2} \Big) R_0 \sin (\Phi_0 - \theta_i) - \frac{K}{\sigma^2} R_1 \sin(\Phi_1 - \theta_i) + \frac{K r_i}{\sigma^2} \Bigg[ \tilde{S}_+ \sin \Big( \Psi_+ - (\phi_i+\theta_i) \Big) - \tilde{S}_- \sin \Big( \Psi_- - (\phi_i- \theta_i) \Big) \Bigg]
\end{align*}
\noindent
where
\begin{align}
H_r(r_i, \phi_i) &=  \frac{1}{N} \sum_{j} \Big( r_j \cos(\phi_j - \phi_i) - r_i \Big) ( 1 - d_{ij}^{-2} )  \label{Hr} \\
H_{\phi}(r_i, \phi_i) &=  \frac{1}{N} \sum_{j} \frac{r_j}{r_i} \sin(\phi_j - \phi_i) ( 1 - d_{ij}^{-2} ), \label{Hphi}  \\
Z_0 = R_0 e^{i \Psi_0}&=  \frac{1}{N} \sum_{j}  e^{i \theta_j}, \label{Qzero} \\
\hat{Z}_0 = \hat{R}_0 e^{i \hat{\Psi}_0}&=  \frac{1}{N} \sum_{j \in N_i}  e^{i \theta_j}, \label{Qzero1} \\
Z_2 = R_2 e^{i \Psi_2}&=  \frac{1}{N} \sum_{j} r_j^2 e^{i \theta_j}, \label{Q1} \\
\hat{Z}_2 = \hat{R}_2 e^{i \hat{\Psi}_2}&=  \frac{1}{N} \sum_{j \in N_i} r_j^2 e^{i \theta_j}, \label{Q2} \\
\tilde{W}_{\pm} = \tilde{S}_{\pm} e^{i \Psi_{\pm}}&=  \frac{1}{N} \sum_{j} r_j e^{i (\phi_j \pm \theta_j)} 
\label{Qplusminus} \\
\hat{W}_{\pm} = \hat{S}_{\pm} e^{i \hat{\Psi}_{\pm}}&=  \frac{1}{N} \sum_{j \in N_i} r_j e^{i (\phi_j \pm \theta_j)}
\end{align}
\noindent
where the $\hat{Z_0}, \dots$ order parameters are summed over all the neighbours $N_i$ of the $i$-th swarmalator: those within a distance $\sigma$. Notice that rainbow order parameters $\tilde{W}$ here are weighted by the radial distance $r_j$. Assuming $\sigma > max(d_{ij})$, we can set $\hat{Z_0} = Z_0, \hat{Z}_1 = Z_1, \hat{W_\pm} = W_{\pm}$. Then $S_{\pm} \sin(\Phi_{\pm} - (\phi \pm \theta) )$ etc of the ring model starting to emerge. If we assume there is no global synchrony $Z_0 = Z_2 = 0$, which happens generically in the frustrated parameter regime $K < 0, J > 0$, and transform to $\xi_i = \phi_i + \theta_i$ and $\eta_i = \phi_i - \theta_i$ coordinates the ring model is revealed 
\begin{align}
\dot{r_i}&= \tilde{\nu}(r_i,\phi_i) +  \frac{J}{2} \Bigg[ \tilde{S}_+ \cos \Big( \Phi_+- \xi_i \Big) + \tilde{S}_{-}\cos \Big( \Phi_{-} - \eta_i \Big) \Bigg] \\
\dot{\xi_i}&= \tilde{\omega}(r_i, \phi_i) + \Bigg[ J_+(r_i) \tilde{S}_+ \sin \Big( \Psi_+ - \xi_i \Big)  + J_- (r_i) \tilde{S}_- \sin \Big( \Psi_{-} -  \eta_i \Big) \Bigg]   \\
\dot{\eta_i} &= \tilde{\omega}(r_i, \phi_i) +  \Bigg[ J_-(r_i) \tilde{S}_+ \sin \Big( \Psi_+ - \xi_i \Big) - J_+(r_i) \tilde{S}_- \sin \Big( \Psi_- - \eta_i \Big) \Bigg]
\end{align}
where
\begin{align}
\tilde{\nu}(r_i, \phi_i) &= H_r(r_i, \phi_i) \\ 
\tilde{\omega}(r_i, \phi_i) &= H_{\phi}(r_i, \phi_i) \\
J_{\pm}(r_i) &= \frac{J}{2 r_i} \pm \frac{K r_i}{\sigma^2}
\end{align}

In the spirit of minimalism, we suppress the $\phi_i$ dependence in the $\nu, \omega$ in our definition of the ring model in the main text. Hence the reported ring model is the  'essence' of the angular piece of the 2D model, as described. 

\end{widetext}

\end{document}


\title{Supplementary Materials}

\maketitle



\section{Generalized OA ansatz: WORK IN PROGRESS}
We write the EOMs as follows
\begin{align}
    \dot{\xi_i} = \omega_i^+ +  \frac{1}{2}\Big[Q_{++} \sin(\Theta_{++} - \xi_i) +  Q_{--} \sin(\Theta_{--} - \eta_i)\Big] \\
    \dot{\eta_i} = \omega_i^- + \frac{1}{2}\Big[Q_{-+} \sin(\Theta_{-+} - \xi_i) +  Q_{+-} \sin(\Theta_{+-} - \eta_i)\Big] 
\end{align}
where
\begin{align}
\omega_i^{\pm} &= \omega_i \pm \nu_i \\
U_{\pm \pm} &= Q_{\pm\pm} e^{i \Theta_{\pm \pm}} =  \frac{1}{N} \sum_j (J_j \pm K_j) e^{i (x_j \pm\theta_j)}
\end{align}
Following \cite{yoon2022sync}, we derive a generalized OA ansatz
\begin{align}
\dot{\alpha} &= -i \omega_+ + \frac{1}{2}( U_{++}^* - U_{++} \alpha^2  ) + \frac{1}{2} \frac{\alpha}{\beta}( U_{+-}^* + U_{+-} \beta^2  ) \\
\dot{\beta} &= -i \omega_- + \frac{1}{2}( U_{--}^* - U_{--} \beta^2  ) + \frac{1}{2} \frac{\beta}{\alpha}( U_{-+}^* + U_{-+} \alpha^2  )
\end{align}

{\color{blue}{
For the model with $J_j=J$ and $K_j=K$, the generalized OA equations are given by 

\begin{align}
\dot{\alpha} &= -i\omega^+ \alpha + \frac{J_+}{2}( W_{+}^* - W_{+}\alpha^2) + \frac{J_{-}}{2}\alpha( W_{-}^* \beta^* - W_{-} \beta ) \\
\dot{\beta} &= -i \omega^- \beta + \frac{J_+}{2}( W_{-}^* - W_{-} \beta^2) + \frac{J_{-}}{2} \beta( W_{+}^* \alpha^* - W_{+} \alpha)
\end{align}
}}

{\color{red}{
For our model with $J_j$ and $K_j$, the OA equations are given by
\begin{align}
\dot{\alpha} &= -i \omega^+ \alpha + \frac{1}{4}( U_{++}^* - U_{++} \alpha^2) + \frac{1}{4}\alpha( U_{--}^* \beta^* - U_{--} \beta ) \\
\dot{\beta} &= -i \omega^- \beta + \frac{1}{4}( U_{+-}^* - U_{+-} \beta^2) + \frac{1}{4} \beta( U_{-+}^* \alpha^* - U_{-+} \alpha)
\end{align}
}}
The mean fields are
\begin{align}
    U_{+, \pm} &= \int (J \pm K) \alpha^*(\nu, \omega, J, K) g(\omega) g(\nu) g(J) g (K) \\
    U_{-, \pm} &= \int (J \pm K) \beta^*(\nu, \omega, J, K) g(\omega) g(\nu) g(J) g (K)
\end{align}
where I have abused notation and used a single $g(.)$ for the distributions of $\nu, \omega, J, K$. 

\begin{itemize}
    \item There are now FOUR mean fields
    \item $\alpha = \alpha(\omega, \nu, J, K, t)$ (same for $\beta$)
    \item Maybe figure out better notation for $U_{\pm, \pm}$ -- Maybe $U_{\pm}^{\pm}$?
\end{itemize}

\section{Movies}

\begin{itemize}
    \item Explanation of movies
\end{itemize}

\bibliography{ref.bib}